\documentclass[12pt]{article}

\usepackage{amsmath,amssymb}
\usepackage{graphicx}

\makeatletter

\renewcommand\section{\@startsection{section}{1}{\z@}
                                   {-3.5ex \@plus -1ex \@minus -.2ex}
                                   {2.3ex \@plus .2ex}
                                   {\normalfont\large\bfseries}}
\renewcommand\subsection{\@startsection{subsection}{2}{\z@}
                                   {-3.25ex\@plus -1ex \@minus -.2ex}
                                   {1.5ex \@plus .2ex}
                                   {\normalfont\normalsize\bfseries}}
\renewcommand\subsubsection{\@startsection{subsubsection}{3}{\z@}
                                   {-3.25ex\@plus -1ex \@minus -.2ex}
                                   {1.5ex \@plus .2ex}
                                   {\normalfont\normalsize\bfseries}}
\renewcommand\paragraph{\@startsection{paragraph}{4}{\z@}
                                   {3.25ex \@plus1ex \@minus.2ex}
                                   {-1em}
                                   {\normalfont\normalsize\bfseries}}
  
\makeatother

\newcommand{\be}{\begin{equation}}
\newcommand{\ee}{\end{equation}}
\newcommand{\bea}{\begin{eqnarray}}
\newcommand{\eea}{\end{eqnarray}}
\newcommand{\ba}{\begin{array}}
\newcommand{\ea}{\end{array}}

\newcommand{\id}{\hbox{1\kern-.27em l}}

\newcommand{\half}{ {\textstyle \frac{1}{2}  } }

\newcommand{\al}{\alpha}
\newcommand{\ga}{\gamma}
\newcommand{\Ga}{\Gamma}
\newcommand{\bet}{\beta}

\newcommand{\ka}{\kappa}
\newcommand{\de}{\delta}

\newcommand{\ep}{\epsilon}
\newcommand{\si}{\sigma}
\newcommand{\la}{\lambda}

\newcommand{\ze}{\zeta}

\newcommand{\La}{\Lambda}

\newcommand{\tha}{\theta}

\newcommand{\Ups}{\Upsilon}

\newcommand{\cT}{\mathcal{T}}

\newcommand{\D}{{\rm d}}
\newcommand{\pa}{\partial}
\newcommand{\rar}{\rightarrow}
\newcommand{\non}{\nonumber}

\newcommand{\SO}{\mathrm{SO}}

\newcommand{\U}{\mathrm{U}}

\newcommand{\ts}{\textstyle}

\begin{document}

\begin{center}

\vspace*{5mm}
{\Large\sf Towards relating the kappa-symmetric and  \\ pure-spinor versions of the supermembrane  }

\vspace*{5mm} {\large Mirela B\u{a}b\u{a}l\^{i}c$^{1,2}$, and Niclas
Wyllard$^{1}$}

\vspace*{5mm}
$^1$ Department of Fundamental Physics\\
Chalmers University of Technology\\
S-412 96 G\"oteborg, Sweden\\[3mm]

$^2$ Department of Theoretical Physics\\
University of Craiova\\
200585 Craiova, Rom\^{a}nia\\[3mm]

{\tt babalac@chalmers.se, wyllard@chalmers.se}

\vspace*{5mm}{\bf Abstract}

\end{center}

\noindent We study the relation between the $\ka$-symmetric
formulation of the supermembrane in eleven dimensions and 
the pure-spinor version.  Recently, Berkovits related 
the Green-Schwarz and pure-spinor superstrings.
In this paper, we attempt to extend  this method to the supermembrane.
 We show that it
is possible to reinstate the reparameterisation constraints in the
pure-spinor formulation of the supermembrane 
by introducing a topological sector and
performing a similarity transformation. The resulting BRST charge is
then of conventional type and is argued to be (related to) the BRST
charge of the $\ka$-symmetric supermembrane in a formulation 
 where all second class constraints 
are `gauge unfixed' to first class constraints. 
In our analysis we also encounter a natural candidate
for a (non-covariant) supermembrane analogue of the superstring $b$ ghost.

\setcounter{equation}{0}
\section{Introduction}
The pure spinor formulation of the superstring \cite{Berkovits:2000a} has 
proven to be quite useful for quantising the superstring in a manifestly 
super-Poincar\'e covariant manner. At first, the `origin' of the formalism 
and its relation to the Green-Schwarz and Ramond-Neveu-Schwarz formulations 
was very mysterious. This state of affairs has gradually improved 
(see e.g.~\cite{Berkovits:2001a,Grassi:2001,Aisaka:2002,Chesterman:2002,Matone:2002,Berkovits:2004f,Aisaka:2005,Berkovits:2007}) and by now the relation to the other 
versions of the superstring is better understood (although some aspects are not yet 
completely satisfactory).

Recently, the relation between the ($\ka$-symmetric) Green-Schwarz
superstring \cite{Green:1983} and the pure spinor version was
clarified~\cite{Berkovits:2007}. The Green-Schwarz formulation of
the open superstring (or one of the two sectors of the closed
superstring) has one reparameterisation constraint, $T$, and 16
fermionic constraints, $d_\al$, half of which are second class and the
other half first class. The separation of the two types of constraints in a
Lorentz-covariant manner, preserving the full ten-dimensional
symmetry, is not possible. Giving up manifest covariance, the usual 
way to treat the $d_\al$ constraints starts by  constructing the Dirac
bracket from the second class constraints. Instead of this approach,
an alternative is to try to view the eight second class constraints
as four first class constraints plus four gauge fixing conditions.
This ``gauge unfixing''  method is not very well known and it is not
known if it can always be applied (see
e.g.~\cite{Harada:1988,Mitra:1990,Vytheeswaran:2000}). In the case of
the superstring however, the method can be
used~\cite{Berkovits:2007} and the resulting set of twelve fermionic
first class constraints can then, together with $T$, be used to write a
conventional BRST charge. 
This BRST charge is not manifestly 
Lorentz covariant. However, after a similarity transformation it 
can be shown to be equal to the pure spinor BRST charge
plus a topological term which decouples due to the quartet
mechanism~\cite{Berkovits:2007}\footnote{The quartet may
actually not decouple completely in all sectors of the theory, see
\cite{Berkovits:2007} for more details.}. 

After the decoupling, the resulting 
BRST charge agrees with the pure spinor BRST charge thereby 
establishing the equivalence between the two formalisms.
The decoupling of the topological quartet effectively
removes the reparameterisation constraint together with one of the
fermionic constraints and the corresponding ghosts, and reinstates Lorentz covariance. 
The remaining eleven bosonic ghosts build up a pure spinor --- eleven being the dimension 
of such a spinor in ten dimensions.

In this paper we attempt to extend the method of
\cite{Berkovits:2007} to the case of the supermembrane in eleven dimensions. 
There are two formulations of this object: a $\ka$-symmetric version \cite{Bergshoeff:1987a}  (analogous to the Green-Schwarz formulation for superstrings) and 
a pure spinor formulation \cite{Berkovits:2002a}.
The supermembrane models are much more involved than the corresponding superstring models, essentially because of the non-linear nature of the world volume theories. Nevertheless, progress can be made.

As a warm-up exercise
and to fix our notation we treat the superparticle in eleven dimensions in the next
section. Then in section \ref{m2} we tackle the much more
complicated supermembrane case.
 In the appendix we collect our conventions and some technical details.

\setcounter{equation}{0}
\section{Superparticle in eleven dimensions}
In this section we  discuss the superparticle in eleven dimensions.
We show that the method of \cite{Berkovits:2007} goes through
essentially unchanged for this case. The superparticle provides a
useful stepping stone towards the much more difficult supermembrane
which we treat in section \ref{m2}.

\subsection{The $\ka$-symmetric superparticle}
The superparticle has the following action \cite{Brink:1981} 
\be \label{Sk}
S =
\int \D \tau (P_M \Pi^M - \half e P_M P^M) =\int \D
\tau  \, \frac{1}{2e} \Pi_M \Pi^M \,,
\ee 
where $e$ is the einbein, $\Pi^M = \dot{X}^M
- i \tha \Ga^M \dot{\tha}$ with $M=0,\ldots,9,11$,  $\tha^A$  is a 32
component Majorana spinor, and $\dot{}\equiv \frac{\pa}{\pa \tau}$ (see the appendix 
for more details on our conventions). 
The conjugate momenta to $X^M$ and $\tha^A$ are denoted $P_M$ and $p_A$, 
respectively. The action (\ref{Sk}) is invariant under the (global) 
supersymmetry transformations
\be 
\de \tha^A = \ep^A \,,\qquad \de x^M = i (\ep \Ga^M \tha)\,,\qquad  \de e=0=\de P_M\,,
\ee 
as well as under the following local fermionic symmetry (`$\ka$-symmetry') \cite{Siegel:1983}:
\be 
\de \tha^A =  P_M (\Ga^M\ka)^A \,,\qquad \de x^M = i (\tha \Ga^M  \de \tha)\,, \qquad \de P_M = 0 \,, \qquad \de e = 4i (\dot{\tha} \ka) \,.
\ee 

From the usual Dirac analysis one obtains the constraints 
\bea
T &=& P_M P^M \approx 0 \,, \non \\
d_A &=& p_A - i P_M (\Ga^M \tha)_A \approx 0\,.
\eea
Here $T$ is the reparameterisation constraint.
As is well-known, the 32 fermionic
$d_A$ constraints comprise 16 first class and 16 second class
constraints.

The basic Poisson brackets are 
\be 
\{P_M, X^N\} = -\de^N_M\,, \qquad
\{p_A,\tha^B\} = -\de_A^B\,. \ee 
From these results it follows that the non-vanishing Poisson brackets 
involving  $\Pi_M$ and $d_A$ are 
\be 
\{d_A , d_B\} = 2i\Ga^M_{AB} P_M\,, \qquad \{d_A,\Pi^M\} = i (\Ga^M\dot{\tha})_A \,.
\ee

\subsection{The pure spinor superparticle}
The pure spinor version of the superparticle in eleven dimensions
was proposed in~\cite{Berkovits:2002a} (see also
\cite{Anguelova:2004}). The action is 
\be \label{Sp}
S = \int \D \tau (P_M
\dot{X}^M - p_A \dot{\tha}^A + w_A \dot{\la}^A - \half P_M P^M) \,.
\ee 
Here the bosonic (i.e.~Grassmann even) spinor $\la^A$ is a pure spinor, 
i.e.~it satisfies $\la \Ga^M \la =0$. Such a spinor has 23 independent 
components (see the appendix for an explicit demonstration of this fact).
The canonical momentum to $\la^A$, denoted $w_A$, therefore has the gauge invariance $\de w_A = \La_M (\Ga^M\la)_A$ induced by the constraint imposed on $\la^A$. This means that $w_A$ and $\la^A$ do not have a canonical (Lorentz covariant) Poisson bracket. 
However, from $w$ and $\la$ one can form gauge-invariant Lorentz-covariant objects, e.g.
\be \label{JNMN}
J = w \la \,, \qquad N^{MN} = \half (w\Ga^{MN}\la) \,,
\ee
which correspond to the $\la$ (or ghost) charge, and the Lorentz current in the $(w,\la)$ sector, respectively. 
For calculations involving such gauge invariant expressions, one can effectively use the canonical Poisson bracket $\{ w_A,\la^B\} = -\de^B_A$, as the non-covariant pieces cancel. 

The BRST charge of the pure spinor model is $Q=\la^A d_A$ and satisfies $\{Q,Q\}=0$.

\subsection{Relation between the two formulations}

We now discuss the relation between the above two formulations for
the superparticle (see \cite{Anguelova:2004} for an alternative,
less direct, approach). In preparation for the supermembrane case we restrict ourselves 
to a classical analysis (i.e.work~at the level
of Poisson brackets). Our discussion closely parallels the
ten-dimensional case discussed in~\cite{Berkovits:2007}.

The first step is to add a topological quartet $(b,c,\beta,\ga)$ to the pure
spinor BRST charge as $Q\rar Q' = \la^A d_A + b \ga$. Here $b$, $c$ are fermionic and $\bet$, $\ga$ are bosonic. 
The canonical Poisson brackets between the new variables are 
\be
\{b,c\} = -1\,, \qquad \{ \bet,\ga\} = -1 \,.
\ee
If one performs the transformation 
\be 
Q'' = e^{ c R/\ga } Q'e^{-c R/\ga } \equiv Q' + \{\frac{cR}{\ga}, Q'\} +  \frac{1}{2!}  \{\frac{cR}{\ga}, \{\frac{cR}{\ga}, Q'\}\} + \ldots \,,
\ee 
where 
\be \label{Rsp}
R = -\frac{i}{2}P_M (d \Ga^M \xi ) \,,
\ee  
and then uses the result
\be
\{Q,R\} = (\la \xi)T \,,
\ee
with the identification $ \ga = -(\la \xi)$, one finds that 
\be  \label{Q''}
Q'' = \la^A d_A -  R + c T + b \ga\,.
\ee 
In this way, the reparameterisation constraint $T$  has been
introduced into the pure spinor formulation.

The next step is to relate the BRST charge $Q''$ to the BRST charge in the
$\ka$-symmetric version of the superparticle. As discussed in the
introduction this is done by replacing the second class constraints
in the $\ka$-symmetric formulation by first class constraints (which upon gauge fixing would give back
the second class constraints). This can be done as in
\cite{Berkovits:2007} by first splitting the $d_A$ constraints into
two parts (e.g.~using lightcone variables); one containing 16 first
class constraints and one containing 16 second class constraints,
and then `gauge unfixing' the 16 second class constraints into 8
first class constraints. However, rather than following this path
one can try to directly find a suitable set of first class
constraints.

It is clear that $Q= \la^A d_A $ with $\la$ pure corresponds to a set of 23
first class constraints, since $\la^A$ has 23 independent components and $\{Q,Q\}=0$. 
To try to extend this number one can make
the Ansatz $Q_0 =  \la^A d_A + \beta^A d_A$ where $\beta^A$ can
depend on $P_M$, but since $P_MP^M$ is a constraint and
$\{d_A,P_MP^M\}=0$ one can take the dependence to be linear, and
make the Ansatz $\bet^A d_A = \frac{i}{2} P_M (d \ga^M \xi)$. One
finds $\{Q_0,Q_0\} = -2(\la \xi) T - {\ts \frac{i}{2}}  P_M (\xi \Ga^M \xi) T$. To simplify
this result one can require  $\xi \Ga^M \xi =0$. Adding also the
$T$ constraint and its associated $(b,c)$ ghosts one sees that 
\be 
\tilde{Q} =
\la^A d_A + \frac{i}{2} P_M (\xi \Ga^M d) + c T - b (\la\xi) 
\ee
satisfies $\{\tilde{Q},\tilde{Q}\} = 0$ and agrees with the above expression
(\ref{Q''}) provided $\la\xi = -\ga$. An explicit solution to $\xi \Ga^M \xi
=0$ and $\la\xi=-\ga$ can of course be found; e.g.~in the $\U(5)$
basis (cf.~appendix \ref{app}), $\xi^- =  -\ga/\la^+$ with all
the other components of $\xi$ being zero is such a solution.

Above we started from the pure spinor model and arrived at the $\ka$-symmetric model. The argument  can of course just as easily  be run in reverse. However, in the supermembrane case treated in the next section it turns out to be easier to start from the pure spinor model. As in~\cite{Berkovits:2007}, one can also map the two actions (\ref{Sk}) and (\ref{Sp}); we will not repeat the details here.

\setcounter{equation}{0}
\section{Supermembrane in eleven dimensions}\label{m2}
In this section we discuss the extension of the method described above 
to the supermembrane. As expected, the supermembrane case is significantly more involved.

\subsection{The $\ka$-symmetric supermembrane}
A $\ka$-symmetric action for the supermembrane in eleven dimensions
was written down by Bergshoeff, Sezgin and
Townsend~\cite{Bergshoeff:1987a}. In a flat supergravity background
the action is 
\bea
S &=& \int\D\tau\D^2\si  \bigg[P_M \Pi_0^M + e^0(P_MP^M + M) + e^i \Pi_i^M P_M  \non \\
&&- {\ts \frac{i}{2}} \ep^{IJK} (\tha \Ga_{MN} \pa_I \tha)\big[\Pi^M_J \Pi_{K}^{N} + i \Pi^M_J (\tha \Ga^{N} \pa_K \tha)- {\ts \frac{1}{3}}(\tha \Ga^{M} \pa_J \tha)(\tha \Ga^{N}
\pa_K \tha) \big] \bigg] \non \\
&=&  -\half\int\D^3\ze \bigg[ \sqrt{-g}(g^{IJ} \Pi_I^M\Pi_{JM} - 1) \\
&&+ i \ep^{IJK} (\tha \Ga_{MN} \pa_I \tha)\big[\Pi^M_J \Pi_{K}^{N} + i \Pi^M_J
(\tha \Ga^{N} \pa_K \tha)-  {\ts \frac{1}{3}}(\tha \Ga^{M} \pa_J \tha)(\tha \Ga^{N}
\pa_K \tha)\big] \bigg]  \,,\non 
\eea 
where $\ze^I=(\tau,\si^i)$ with $I,J,K =0,1,2$, and  $i,j=1,2$. Also, 
\be
\Pi_I^M = \pa_I X^M -i \tha\Ga^M\pa_I\tha\,,
\ee
$P_M$ is the conjugate momentum to $X^M$, and 
\be \label{M}
M= \det (\Pi^N_i\Pi_{jN}) = \half\ep^{ij} \Pi^M_i \Pi^N _j \ep^{kl} \Pi_{kM} \Pi_{lN} \,.
\ee
The two forms of the action above are related by integrating out $P_M$ and using the parameterisation \cite{Bergshoeff:1987b}
\be
g^{ij} = \ga^{ij} - \frac{N^iN^j}{N^2} \,, \qquad g^{0i} = \frac{N^i}{N^2} \,, \qquad g^{00} = -\frac{1}{N^2} \,,
\ee
together with the identifications
\be
e^0 = \frac{N}{2\sqrt{\ga}} \,, \qquad e^i = -N^i \,, 
\ee 
and the result \cite{Bergshoeff:1987b}
\be
M = \ga (\ga^{ij} \Pi_i^M \Pi_{jM} - 1)\,.
\ee

The above action is invariant under global supersymmetry as well as under a local fermionic $\ka$-symmetry (we do not show this here; see e.g.~\cite{Bergshoeff:1987a,Berkovits:2002a,Aisaka:2006} for details).

Of particular interest for us is the
hamiltonian analysis of the constraints derived from the above action. 
Such an analysis was performed in~\cite{Bergshoeff:1987b}. 
The reparameterisation
constraints are 
\bea \label{Ts}
T &=& K_M K^M + M - 2 \ep^{ij}\Pi_i^M (d \Ga_M \pa_j \tha) \approx 0\,, \non \\
T_i &=& K_M \Pi_i^M - d  \pa_i \tha \approx 0 \,,
\eea 
where
\be
K_M = P_M - i\ep^{ij}(\tha\Ga_{MN}\pa_i\tha)(\Pi^N_j
 + \ts\frac{i}{2}\tha\Ga^N\pa_j\tha) \,,
\ee
and $M$ is as in (\ref{M}).
The fermionic constraints are 
\bea d_A &=& p_A - iP_M(\Ga^M\tha)_A -
\ts\frac{i}{2}\ep^{ij}(\Ga_{MN}\tha)_A \bigg[\Pi^M_i \Pi^N_j + i\Pi^M_i
(\tha\Ga^N\pa_j\tha) \\&&-
\ts\frac{1}{3}(\tha\Ga^M\pa_i\tha)(\tha\Ga^N\pa_j\tha) \bigg]  -
\ts\frac{1}{2}\ep^{ij}(\tha\Ga_{MN}\pa_i\tha)(\Ga^M\tha)_A [\Pi^N_j
 + \ts\frac{2i}{3}\tha\Ga^N\pa_j\tha ] \approx 0\,.\non 
\eea 

The basic canonical Poisson brackets are
\be
\{ P_M(\si), X^N(\rho)\} = -\de_M^N\de^2(\si-\rho) \,, \qquad \{p_A(\si),\tha^B(\rho)\} = -\de_A^B\de^2(\si-\rho)\,,
\ee
which imply the following non-vanishing Poisson brackets between 
$K_M$, $\Pi_i^M$ and $d_A$:
\bea
{} \{d_A(\si),d_B(\rho) \} &=& 2i K_M \Ga^M_{AB} \, \de^2(\si-\rho) + i\ep^{ij}\Pi_{iM} \Pi_{jN} \Ga^{MN}_{AB} \, \de^2(\si-\rho) \,,  \non\\
{} \{d_A(\si), K_M(\rho)\} &=& -2i \ep^{ij}\Pi^N_{i} (\Ga_{MN}\pa_j\tha)_A \,\de^2(\si-\rho) \,, \non\\
{} \{d_A(\si),\Pi_i^M(\rho)\} &=& 2i (\Ga^M \pa_i \tha)_A \, \de^2(\si-\rho) \,, \\
{} \{K_M(\si),K_N(\rho) \} &=& -i\ep^{ij}(\pa_i \tha \Ga_{MN} \pa_j \tha)\,\de^2(\si-\rho) \,, \non \\
{} \{K_M(\si),\Pi^N_i(\rho) \} &=& -\de^N_M\frac{\pa}{\pa \rho_i}\de^2(\si-\rho) \,. \non
\eea
Here, all fields on the right hand side depend on $\rho$. 
To obtain these results it is important to keep track of the dependent variable of the fields, writing $\Ups(\si) = \Ups(\rho + (\si-\rho))$ for any field that depends on $\si$ and Taylor expanding, as well as making use of the relation $x\pa_x \de(x) = -\de(x)$.

\subsection{The pure spinor supermembrane}
A pure spinor version of the supermembrane in eleven dimensions was
proposed by Berkovits in ref.~\cite{Berkovits:2002a}. This model is based on
the action (in our conventions)
\bea 
\label{SM2p} S &=& \int \D\tau\D^2\si \bigg[ K_M \Pi_0^M - d \pa_0 \tha + w \pa_0 \la \non \\ &&-{\ts \frac{i}{2}} \ep^{IJK} (\tha \Ga_{MN} \pa_I \tha)(\Pi^M_J \Pi^N_{K} + i\Pi^M_J(\tha \Ga^{N} \pa_K \tha)- {\ts \frac{1}{3}}(\tha \Ga^{M} \pa_J \tha)(\tha \Ga^{N} \pa_K \tha) \non \\
&& -\half \big[ K_M K^M + M + 2\ep^{ij} (d \Ga_M \pa_i \tha) \Pi_j^M + 2\ep^{ij} (w \Ga_M \pa_i \la) \Pi_j^M  \non \\
&&\quad \;\; +4i \ep^{ij} (w \Ga^M \pa_i\tha)(\la\Ga_M\pa_j\tha) - 4i \ep^{ij} (w\pa_i\tha)(\la \pa_j \tha) \big] \\
&&\qquad +\, e^i \big[ K_M \Pi^M_i - d\pa_i\tha + w\pa_i\la \big] \bigg] .\non
\eea 
Note that this action reduces to that of the superparticle by throwing away all dependence on $\si^i$. Analogously to the superparticle case, the proposed BRST charge is
\be
Q = \int \D^2 \si \la^A d_A\,,
\ee  
where $\la^A$ satisfies the pure spinor constraint $\la \Ga^M \la =0$. 
However, in contrast to the superparticle case,  further constraints are needed to make $Q$ nilpotent and the action BRST invariant. 
As shown in \cite{Berkovits:2002a} the following constraints 
also seem to be required 
\be  \label{addlconstr}
\qquad (\la \Ga_{MN}
\la) \Pi_i^N = 0\,, \quad \la \pa_i \la=0 \,.
\ee 
These constraints are more puzzling and appear to be at a different level from the pure spinor constraint. Note that the constraints (\ref{addlconstr}) are BRST closed. 

In a more recent development, a lagrangian approach was taken which leads to the same
constraints \cite{Fre:2006} (see also~\cite{Fre:2008}). This analysis was performed essentially without making any a priori assumptions, which lends additional support to the constraints (\ref{addlconstr}). Still, the exact form of the full set of constraints deserves further study.

We should also mention another attempt to understand the origin of
the pure spinor supermembrane \cite{Aisaka:2006}. In this paper the
goal was to derive the pure spinor model starting from a ``doubled''
version of the $\ka$-symmetric supermembrane. This approach was
partially successful, but was not as complete as that for the
superstring \cite{Aisaka:2005}, due to the intricate nonlinear
nature of the supermembrane.

\subsection{Relation between the two formulations}
A natural first step to relate the above two formulations is to try to find a supermembrane generalisation of the superparticle result (\ref{Rsp}). 
We propose that the following expression provides such a generalisation
\bea \label{R}
R &=& \int\D^2\si \big[-\ts\frac{i}{2}K_M (d \Ga^M \xi) + \ts\frac{i}{4}\ep^{ij} \Pi_i^M\Pi_j^N (d \Ga_{MN} \xi) \non \\
&&-\, \half  \ep^{ij} \Pi_i^M (\xi \Ga_M \pa_j \tha) (w \lambda)
 - {\ts \frac{1}{4}}  \ep^{ij} \Pi_i^M (\xi \Ga_{MNR}  \pa_j \tha) (w \Ga^{NR} \lambda)  \\
&&-\, \half  \ep^{ij} \Pi_i^M (\xi  \pa_j\tha) (w \Ga_{M} \lambda) -
 {\ts \frac{1}{4}}  \ep^{ij} \Pi_i^M (\xi \Ga^{NR}  \pa_j\tha) (w \Ga_{MNR}
\lambda) \big] \,.\non 
\eea
A few comments about this expression are in order. 
The second line of this expression is invariant
under the gauge transformation $\de w_A = \La_M (\Ga^M \la)_A$
arising from the fact that $\la$ is pure, i.e.~$\la \Ga^M \la=0$. This is easy to see since it involves the gauge invariant expressions encountered previously in the superparticle case (\ref{JNMN}). 
The first term on the third line is not invariant unless 
one imposes additional conditions. Provided that $(\la \Ga_{MN} \la) \Pi^M_i =0$ 
it is invariant. However, even with this additional condition the 
second term on the third line of (\ref{R}) is not invariant; instead its variation becomes proportional to $ \La_P \Pi^P_i \ep^{ij} (\xi \Ga_{MN} \pa_j\tha) (\la\Ga^{MN} \la)$. If one imposes the stronger condition $\la\Ga^{MN} \la=0$ (or  the slightly weaker condition ($\la \Ga^{MN}\la)\Pi^P_i=0$), it is invariant, but this possibility appears disfavoured since in previous work \cite{Berkovits:2002a,Fre:2006} constraints stronger than (\ref{addlconstr}) were not necessary. 
Another possibility is that $\{Q,\de R\}=0$, i.e.~$R$ is only gauge invariant up to BRST closed terms. Although some terms in  $\{Q,\de R\}$ can be cancelled if one also imposes $\la \pa_i \la=0$, it seems that not all terms can be made to vanish, even using Fierz identities. Therefore we are left with a puzzle regarding the final term in (\ref{R}). For the remainder of this paper we will assume that either the stronger condition $\la \Ga^{MN}\la=0$ can be used in our calculations, or that there is another way to make the final term in (\ref{R}) gauge invariant so that the non-covariant pieces in the bracket with $Q$ vanish. Note that the first possibility is not in conflict with the superparticle result since in our calculations $\la \Ga^{MN}\la$ is always multiplied by expressions that vanish in the superparticle limit.

A strong argument in favour of (\ref{R}) is related to its behaviour under the double
dimensional reduction to the $d=10$ type IIA superstring case.
Under this reduction one has
\be \label{d10conds}
\Pi_2^M = \de_{11}^M \,, \qquad \pa_2 \tha = 0 \,, \qquad K_{11} = \La_{11}= 0 \,.
\ee
It is easy to check that when these conditions are fulfilled, $R$ as written above is gauge invariant. Furthermore,  by implementing the conditions (\ref{d10conds}) into $R$ leads to 
\bea
R_{d=10} &=& \int \D \si_1 \bigg[-\ts\frac{i}{2}K_m (d \Ga^m \xi) + \ts\frac{i}{2}\Pi_1^m (d \Ga_{m}\Ga^{11} \xi) \non \\
&&-\, \half   (\xi \Ga^{11}  \pa_1 \tha) (w \lambda)
 - {\ts \frac{1}{4} }   (\xi \Ga^{11}\Ga_{nr} \pa_1 \tha) (w \Ga^{nr} \lambda)  \\
&&-\, \half   (\xi  \pa_1\tha) (w \Ga^{11} \lambda)  - {\ts \frac{1}{4} }   (\xi
\Ga_{nr}  \pa_1\tha) (w \Ga^{11}\Ga^{nr} \lambda)\bigg] \non ,
\eea 
where
$m,n,r=0,\ldots,9$. Splitting $\xi^A
=(\xi^\al,\tilde{\xi}^{\dot{\al}})$ and similarly for $d_A$, $\tha_A$, $w_A$ and $\la^A$, 
we find $R = \xi^\al G_{\al} +
\tilde{\xi}^{\dot{\al}}\tilde{G}_{\dot{\al}}$ where 
\bea
G_\al &=& -\ts\frac{i}{2}K_m ( d \ga^m)_\al + \ts\frac{i}{2} \Pi^m (d \ga_{m} )_\al \non \\
&&-\, \half (\pa \tha)_\al (w \lambda)
 - {\ts \frac{1}{4} }  (\ga_{mn}  \pa \tha)_\al (w \ga^{mn} \lambda)  \,, \non \\
\tilde{G}_{\dot{\al}} &=& -\ts\frac{i}{2}K_m (\tilde{d}\ga^m)_{\dot{\al}} - \ts\frac{i}{2}\Pi^m ( \tilde{d}\ga_m)_{\dot{\al}}  \non \\
&&-\, \half  (\bar{\pa} \tilde{\tha})_{\dot{\al}} (\tilde{w}\tilde{\lambda})
 - {\ts \frac{1}{4} }(\ga_{mn}  \bar{\pa} \tilde{\tha})_{\dot{\al}} (\tilde{w} \ga^{mn} \tilde{\lambda})  \,,
\eea 
which precisely corresponds to the
ten-dimensional result \cite{Berkovits:2001a,Oda:2004}, taking into account differences in conventions (we also used the superstring equations of motion for $\tha$ and $\tilde{\tha}$). 

From (\ref{R})  a lengthy calculation leads to 
\be 
\{ Q,R
\} = \int \D^2\si [(\la \xi) \cT - 2 (\la \Ga_M \xi) \ep^{ij} \Pi_i^M \cT_j] \,,
\ee
where 
\bea \label{MrTs}
\cT &=& K^M K_M + M - 2 \ep^{ij} \Pi_i^M (d \Ga_M \pa_j \tha) - 2 \ep^{ij} \Pi_i^M (w \Ga_M \pa_j \la) \non \\ && -4i  \ep^{ij}(w \pa_i \tha) (\la \pa_j \tha) +4i \ep^{ij} (w \Ga_M \pa_i \tha) (\la\Ga^M \pa_j \tha) \\
 \cT_i &=& K_M \Pi_i^M - d \pa_i  \tha +  w \pa_i \la\,. \non 
\eea 
This is one of the  main results of this paper. Note that the $\cT$'s (\ref{MrTs}) are ghost completions of the $T$'s (\ref{Ts}).

The expressions in (\ref{MrTs}) are precisely the
combinations that appear in the third and fourth, and the fifth lines 
in the action (\ref{SM2p}). This is a good indication that we are 
on the right track, and gives further support to our proposal for $R$.

If one imposes $\la \Ga_M \xi =0$ and $\la \xi =1$, one finds $\{Q,R\}=T$. 
This implies  that $R$ is an eleven-dimensional analogue of the
(non-covariant) superstring $b$ ghost (the superparticle limit of which was
discussed in \cite{Anguelova:2004}.) It is non-covariant in the
sense of the $Y$-formalism \cite{Oda:2004,Oda:2005}. It may be
possible to extend it to a covariant expression along the lines of
\cite{Berkovits:2004,Berkovits:2006}. 

A natural strategy would be to also try to find an $R_i$ such that
$\{Q,R_i\} = T_i + ...$. An attempt based on $R_i = \int \D^2\si \Pi^M_i
(d \Ga_M \xi)  + ...$ fails since one is forced to impose $\la \Ga^M \xi = 0 =
\la \Ga^{MN} \xi$ and $\la\xi \neq 0$. However, when $(\la \Ga^M
\la)=0$, 
\be
(\la\xi)^2 = \frac{3}{2} (\la \Ga^M \xi)(\la \Ga_M \xi) + \frac{1}{4}
(\la \Ga^{MN} \xi)(\la \Ga_{MN} \xi)\,.
\ee
 Thus, there appears to be no such $R_i$.

As in the superparticle case one can perform transformations 
using $e^{c_\xi R_\xi/\ga_\xi}$ where the $\xi$ subscript indicate that we perform several transformations using $R$'s with various different fixed $\xi$'s. 
This gives leading terms in $Q'$ of the form 
\be \label{Q'0}
Q' = \sum_\xi c_\xi T_\xi +\ldots
\ee 
 where $T_\xi$ is a certain combination of $T$ and the eleven $T^M\equiv \ep^{ij}\Pi_i^M T_j$. 

Possibly the most natural approach would be to pick $T$ and two fixed $M$'s, $\pm$ say, $T^{\pm} = \ep^{ij} \Pi^{\pm}_i T_j$,  so that
\be \label{Q'1}
Q' = cT + c_+ T^+ + c_- T^- + \ldots\,.
\ee
 Although not covariant and not based on the usual form of reparameterisation constraints (\ref{Ts}) this would be part of a viable form for a BRST charge arising from the $\ka$-symmetric formulation. If one insists on covariance one could keep all the $T^M$ so that  
\be  \label{Q'2}
Q' = cT + c_M T^M+\ldots \,.
\ee
In this case the constraints would be reducible, but it may be profitable to keep covariance i.e.~to work with $T^M = \ep^{ij}\Pi_i^M T_j$. It is easy to check that, generically, the two sets of constraints based on $T^M$ or $T_i$ define the same constraint surface. 

Reducible constraints satisfy certain relations between them (see \cite{Henneaux:1994} for a detailed exposition). To find these for the constraints at hand we closely follow the analysis in~\cite{Aisaka:2006}.  
(The set of reducible constraints in that work are not quite the same as ours; it may be possible to find a closer link between the two sets.)

To find the first order reducibility functions $Z^M_p$ where $p=1,\ldots,9$ we want to solve
\be
Z^p_M T^M = 0\,.
\ee
Since $T^M = \ep^{ij} \Pi_{i}^M T_j$ the above equation can be written as $\ep^{ij} Y^p_i T_j = 0$ with $Y^p_i = Z^p_M\Pi^M_i$. The solution is $Y_i^p=C^p T_i$ where $C^p$ can be put equal to 1 by rescaling $Z^p$. In other words, we need
\be
Z_p \cdot \Pi_i = T_i\,.
\ee
This is solved by $Z_p = X_p + W$ where $X_p$ is a nine-vector orthogonal to the plane spanned by $\Pi_i$ and $W$ is a solution to the above equations lying in the $\Pi_i$ plane, i.e.~$W = a^i \Pi_i$. Now, $M_{ij} a^j = T_i$ where $M_{ij}=\Pi_i\cdot \Pi_j$ so the solution is $a^i = (M^{-1})^{ij} T_j$. As in \cite{Aisaka:2006} one can easily show that the reducibility is first order: $c^pZ_p = 0$ implies $c_p=0$.

Thus, we seem to be close to finding a $Q'$ which can be related to a BRST charge in the $\ka$-symmetric formulation, either of the form (\ref{Q'1}) or (\ref{Q'2}). 

However, perhaps somewhat surprisingly we have not been able to find a solution to the condition $\la\xi =0$ and $\la \Ga^M \xi = \de^M_N$ for a fixed $N$ that would be required for this approach to work. If one only imposes $\la \Ga^M \la =0$ it is almost possible, but putting nine components of  $\la \Ga^M \xi$ to zero, the solutions we have found automatically puts the rest of $\la \Ga^M \xi$ to zero (but not $\la\xi$). If one also imposes $\la \Ga^{MN}\la=0$ the situation is worse: in our solutions setting five components of  $\la \Ga^M \xi$ to zero puts the remaining components to zero and also forces $\la\xi$ to be zero. 

The equations one needs to solve are rather complicated (see appendix \ref{app}) so it is possible that there are solutions that can give (\ref{Q'1}) or (\ref{Q'2}), but even if this is not the case, one can use other $\xi$'s and obtain a more general form (\ref{Q'0})  where the $T_\xi$'s are more complicated expressions obtained from a parameterisation of independent ``components'' of  $\xi$. 
It seems that it can still be viewed as the leading part of a viable form of the BRST charge in the $\ka$-symmetric model, but it is far from the most natural choice. This point should be studied further. Also, we only calculated the lowest order terms in the similarity transformation. Although general theorems seem to guarantee that the construction will work also at higher orders since we started from a BRST charge that satisfies $\{Q,Q\}=0$, it may be profitable to work out the details.

Above we only studied how the BRST charges are related. In the same way as in \cite{Berkovits:2007} it should also be possible to relate the two actions. Although our work supports the pure spinor formulation it does not really clarify what constraints should be imposed on $\la$. Partly this is a consequence of the fact that we started from the pure-spinor formulation and tried to obtain the $\ka$-symmetric formulation rather than the other way around. It may be fruitful to start from the $\ka$-symmetric formulation and try to obtain the pure-spinor model. However, as we have seen it appears that in order to obtain the pure-spinor model one should not use the canonical form of the constraints.  

\section*{Acknowledgements}
MB was supported by the EU grant MRTN-CT-2004-512194. 
NW  was supported by a grant from the Swedish Science Council.
We would like to thank Martin Cederwall for a useful conversation. 
We found the GAMMA package \cite{Gran:2001} to be very helpful in handling the large amount of gamma matrix algebra needed in the calculations.

\appendix

\setcounter{equation}{0}
\section{Conventions and technical details} \label{app}
In this appendix we collect our conventions and some technical
details. Our conventions are closely related to those of
\cite{Aisaka:2006}, but with some minor differences.

Spacetime indices are labeled by capital letters from the middle of
the alphabet: $M,N,\ldots = 0,\ldots,9,11$. Spinor indices are
labeled by capital letters from the beginning of the alphabet:
$A,B,\ldots = 1,\ldots,32$. The gamma matrices $(\Ga^M)^A{}_B$
satisfy the usual algebra: $\{ \Ga^M ,\Ga^N \} =
\ts\frac{1}{2}(\Ga^M\Ga^N + \Ga^N\Ga^M)=\eta^{MN}$. Indices
can be lowered using $C_{AB}=-C_{BA}$ via 
$(\Ga^{M})_{AB} = C_{AD} (\Ga^{M})^{D}{}_B$. 
We do not write $C_{AB}$ explicitly as
the position of the indices should always be clear from the context.
Also, we do not write the spinor indices explicitly in fully contracted expressions. 
$\Ga^{M_1\cdots M_p}$ is antisymmetric for $p=0,3,4$ and symmetric
for $p=2,3,5$; these form a basis for the bispinor $\Psi_A \Ups_B$
as 
\bea 
\Psi_A \Ups_B &=& \frac{1}{32}\bigg[(\Psi \Ups) C_{AB} +
(\Psi\Ga^{S_1} \Ups) (\Ga_{S_1})_{AB}- \frac{1}{2!}(\Psi\Ga^{S_1S_2}
\Ups)(\Ga_{S_1S_2})_{AB}-\non\\&&-\frac{1}{3!}(\Psi\Ga^{S_1S_2S_3}
\Ups)(\Ga_{S_1S_2S_3})_{AB}+ \frac{1}{4!}(\Psi\Ga^{S_1S_2S_3S_4}
\Ups)(\Ga_{S_1S_2S_3S_4})_{AB}+\non\\&&+
\frac{1}{5!}(\Psi\Ga^{S_1S_2S_3S_4S_5}
\Ups)(\Ga_{S_1S_2S_3S_4S_5})_{AB}\bigg].
\eea

We sometimes find it useful to decompose our expressions into a 
(non-covariant) $\U(5)$ basis. Alternative decompositions are
$\SO(8)$ and $\SO(9)$. Under $\SO(11)\rar  \U(5)$ a vector
decomposes as $V^M \rar (v^a,v_a,v^{11})$ where 
\be 
v_a = \frac{V^a
+ i V^{a+5}}{2} \,, \qquad  v^a = \frac{V^a - i V^{a+5}}{2} \,,
\qquad v^{11} = V^{11} \,.
\ee 
From which it follows that e.g.~$U_M V^M = 2u_a v^a +
2u^a v_a + u^{11} v^{11}$. Tensors are decomposed in a similar way. 

A spinor
$\Psi^A$ splits as $(\psi_{\al},\psi_{\dot{\al}})$ and then further
as $\psi_\al \rar (\psi^+,\psi^a,\psi_{[ab]})$ and $\psi_{\dot{\al}}
\rar (\psi^-,\psi_a,\psi^{[ab]})$ where $a,b=1,\ldots,5$.

In the $U(5)$ basis the gamma matrices can be chosen as
\be
\ba{rcllll}
{(\ga^1)_A}^B &=& \frac{\si^1+i\si^2}{2} \otimes \id \otimes \id \otimes
\id \otimes \id \,,&
 {(\ga^2)_A}^B &\!\!=&\!\!
\si^3\otimes \frac{\si^1+i\si^2}{2} \otimes \id \otimes \id \otimes \id \,,\\
{(\ga^3)_A}^B &=&
\si^3\otimes \si^3 \otimes\frac{\si^1+i\si^2}{2}  \otimes \id \otimes \id\,,&
{(\ga^4)_A}^B &\!\!=&\!\!
\si^3\otimes \si^3\otimes \si^3 \otimes\frac{\si^1+i\si^2}{2} \otimes \id\,,\\
{(\ga^5)_A}^B &=&
\si^3\otimes \si^3\otimes \si^3 \otimes \si^3 \otimes\frac{\si^1+i\si^2}{2} \,, & C_{AB}&\!\!=&\!\! i\si^2 \otimes\si^1\otimes i\si^2 \otimes\si^1 \otimes i\si^2 \,,\\
{(\ga_1)_A}^B &=& \frac{\si^1-i\si^2}{2} \otimes \id \otimes  \id \otimes
\id \otimes \id\,,&
 {(\ga_2)_A}^B &\!\!=&\!\!
\si^3\otimes \frac{\si^1-i\si^2}{2} \otimes \id \otimes \id \otimes \id \,,\\
{(\ga_3)_A}^B &=&
\si^3\otimes \si^3 \otimes\frac{\si^1-i\si^2}{2}  \otimes \id \otimes \id \,,&
{(\ga_4)_A}^B &\!\!=&\!\! \si^3\otimes \si^3\otimes \si^3
\otimes\frac{\si^1-i\si^2}{2} \otimes \id \,,\\{(\ga_5)_A}^B &=& \si^3\otimes \si^3\otimes \si^3\otimes \si^3
\otimes\frac{\si^1-i\si^2}{2} \,,&
 {(\Ga^{11})_A}^B &\!\!=&\!\!\si^3 \otimes
\si^3\otimes \si^3 \otimes\si^3 \otimes\si^3\,,
\ea
\ee
  where $\si^{1,2,3}$ are the usual Pauli matrices 
\be
 \si^1 =\left(\ba{cc}0&1\\1&0\ea\right) \,,\quad \si^2=\left(\ba{cc}0&-i\\i&0\ea\right)
\,,\quad \si^3=\left(\ba{cc}1&0\\0&-1\ea\right) .\ee 
 Using the  $U(5)$
decomposition, we can write formul\ae{} for $\la\xi$,  $\la \Ga^M\xi $ and
$\la\Ga^{MN}\xi$  in the following form
 \bea
\label{u5}
\la \xi &=& \la^+\xi^- - \la^-\xi^+ +\la^a\xi_a - \la_a\xi^a + \half \la_{ab}\xi^{ab} - \half \la^{ab}\xi_{ab}\non \\
\la \Ga^{11} \xi &=& \la^+\xi^- + \la^-\xi^+ +\la^a\xi_a + \la_a\xi^a + \half \la_{ab}\xi^{ab} + \half \la^{ab}\xi_{ab}\non \\
\la \ga^{a} \xi &=& -\la^+ \xi^a - \la^a\xi^+ + \la_b\xi^{ab}+\la^{ab}\xi_b + \ts\frac{1}{4} \ep^{abcde}\la_{bc}\xi_{de}\non \\
\la \ga_{a} \xi &=& \la^- \xi_a + \la_a\xi^- - \la^b\xi_{ab}-\la_{ab}\xi^b - \ts\frac{1}{4} \ep_{abcde}\la^{bc}\xi^{de} \non \\
\la \ga^{a}\Ga^{11} \xi &=& \la^+ \xi^a + \la^a\xi^+ + \la_b\xi^{ab} + \la^{ab}\xi_b - \ts\frac{1}{4} \ep^{abcde}\la_{bc}\xi_{de} \non \\
\la \ga_a\Ga^{11} \xi &=& \la^- \xi_a + \la_a\xi^- + \la^b\xi_{ab} + \la_{ab}\xi^b - \ts\frac{1}{4} \ep_{abcde}\la^{bc}\xi^{de}  \\
\la \ga^{ab} \xi &=& -\la^+ \xi^{ab} - \la^{ab}\xi^+ - \half \ep^{abcde}(\la_{cd}\xi_e + \la_e \xi_{cd})\non \\
\la \ga_{ab} \xi &=& \la^- \xi_{ab} + \la_{ab}\xi^- + \half \ep_{abcde}(\la^{cd}\xi^e + \la^e\xi^{cd})\non \\
\la \ga^{a}{}_b \xi &=& \la^a\xi_b + \la_b\xi^a + \la_{bc}\xi^{ca} + \la^{ac}\xi_{cb} \non\\
&& \!\! +\, \half \de^a _b(\la^+\xi^- + \la^-\xi^+ - \la^c\xi_c - \la_c\xi^c + \half\la_{cd}\xi^{cd} + \half \la^{cd}\xi_{cd})\non
\eea
 Sometimes we find it useful to decompose further into
$\U(4)$. In this case we write 
\be \label{u4}
\ba{rclcrcl}
\la^a &\rar& (\la^{a'}, \ka^+)\,, &
\la_a &\rar& (\la_{a'}, \ka^-)\,, \\
\la^{ab} &\rar& (\la^{a'b'},\ka^{a'})\,, &
\la_{ab} &\rar& (\la_{a'b'},\ka_{a'})\,, 
\ea
\ee
and similarly for $\xi^A$.
The
corresponding formul\ae{} for $\la \xi$, $\la \Ga^M\xi $ and $\la \Ga^{MN}\xi$
in the $\U(4)$ basis can be obtained from (\ref{u5}) by inserting the
expressions (\ref{u4}) (we will not write the result explicitly).
To simplify the notation, below we drop the prime 
 and use $a,b=1,\ldots,4$. 

For example, in the $\U(4)$ basis one can write explicit solutions
to the $\la \Ga^M\la =0 $ constraint, e.g.
\bea
\la^a &=&\frac{1}{ \la^+ \ka^e \la_e }  \bigg[ \ka^- \ka^a \ka^b\la_b + \la^{ab} \la_b \ka^c\la_c -\la^+ \la^{ab}\la_{bc}\ka^c -{\ts \frac{1}{8}}\ka^a \ep^{bcde}\la_{bc}\la_{de} \non \\ &&\quad  -\half \ka^a \la^+ \la^{bc}\la_{bc} +  {\ts \frac{1}{8} }\la^{ab}\la_{b} \ep^{cdef}\la_{cd}\la_{ef}  -  {\ts \frac{1}{2} } \la^+\la^-\ep^{abcd}\la_{b}\la_{cd}   \bigg],  \non \\
\ka^a &=&  -\frac{1}{ \ka^e\la_e }\bigg[ \la^-\la^+ \la_a - \ka^- \la_{ab}\ka^b - \la_{ab}\la^{bc}\la_c  -\half\la^+ \la_{abcd}\ka^b\la^{cd}   \bigg],  \\
\ka^+ &=&\frac{\ep^{abcd}\la_{ab}\la_{cd} - 8 \ka^a \la_a}{8\la^+} \,,\non
\eea
which shows that $\la$ has 23 independent components. 
It is also possible to write down explicit solutions to the $\la \Ga^M\la =0 =\la
\Ga^{MN}\la$ constraints, e.g. 
\bea
\la^a = \frac{1}{2\la^+}\ep^{abcd}\ka_b \la_{cd} \,, \qquad \ka^+=  \frac{1}{8\la^+}\ep^{abcd}\la_{ab} \la_{cd} \,,\non \\
\la_a = \frac{1}{2\la^-}\ep_{abcd}\ka^b \la^{cd} \,, \qquad \ka^-=  \frac{1}{8\la^-}\ep_{abcd}\la^{ab} \la^{cd}  \,,\\
\la^{ab} = 2 \frac{ \ka^{[a} \ep^{b]cde} \ka_c \la_{de} + 2 \la^+\la^- \ep^{abcd}\la_{cd} }{  \ep^{fghk} \la_{fg}\la_{hk} } \,, \non
\eea
which shows that such a $\la$ has 16 independent components.

\begingroup\raggedright\endgroup

\end{document}